# A UX 3.0 Paradigm Framework: Designing for Human-Centered AI Experiences

Wei Xu, Zhejiang University

## Insights

- User experience practices have evolved in stages and are entering a transformative phase: UX 3.0, driven by AI technologies and shifting user needs.
- Human-centered AI (HCAI) experiences are emerging, necessitating new UX approaches to support UX practices in the AI era.
- I propose a UX 3.0 paradigm framework to respond and guide UX practices in developing HCAI systems.

User experience (UX) refers to the overall quality of a person's interaction with a system, product, or service, encompassing factors such as usability, accessibility, satisfaction, and emotional response throughout the entire user journey. Over the past 40 years, beginning in the early computer era with usability engineering, UX practice expanded during the Internet and mobile revolutions, and is now entering a new stage in the AI era.

The rise of AI-based systems (e.g., generative AI, agentic AI, embodied AI, applications, services) has introduced new challenges and opportunities for UX practice. Human-centered AI (HCAI), a version of a human-centered design (HCD) approach adapted for AI, has emerged to address the risks posed by AI [1,2]. UX design is now recognized as a key factor in developing HCAI systems. However, AI introduces new complexities, such as autonomous features, evolving machine behavior, hallucinations from large language models, output explainability, AI ethics, intelligent user interfaces, and human-AI collaboration, which traditional UX practices have challenges in addressing.

To tackle these challenges and deliver an optimal experience when interacting with AI systems, a UX paradigm tailored to the AI era is lacking. This article responds to that gap by analyzing how UX has evolved across technological eras, identifying current needs, and proposing a new UX 3.0 paradigm for designing HCAI systems. The goal is to guide the advancement of UX practice in the AI era.

## Cross-Era Evolution of UX Practices

Over the past 40 years, UX practice has undergone three main stages, shaped by changes in technology, user needs, and UX approaches. Each of the three stages reflects clear characteristics that represent distinct UX paradigms (Table 1). A UX paradigm serves as a guiding framework that shapes how UX is practiced, including its scope, focus, and methods. Over time, new technologies and evolving user needs have driven the development of these paradigms, advancing the field of UX practice.

| UX Paradigms | **UX 1.0 Paradigm** (exploring stage) (late 1980s – ~ 2007 ) | **UX 2.0 Paradigm** (growing stage) (~ 2007 – ~ 2015 ) | **UX 3.0 Paradigm** (maturing stage) ( ~ 2015 – now) |
|---|---|---|---|
| **Technological platform** | PC / Internet Era | Mobile Internet Era | AI Era |
| **Primary user needs** | Usability | Plus: User experience (beyond UI usability) | Plus: natural interaction, human controllability, explainability, ethics, privacy, decision-making authority, reskilling, etc. |
| **Design philosophy** | Human-centered | Human-centered | Human-centered AI |
| **UX ecosystem** | None | Emerging | Forming |
| **Focus** | Usability of the user interfaces of individual products (siloed approach) | Plus: end-to-end experience | Human-centered AI experience |
| **Scope** | Product development stage | The entire product life cycle | Plus: Organizational, AI ecosystem, sociotechnical environments (beyond UX of individual products) |
| **Methodology** | Usability engineering: UI prototyping, usability testing, etc. | UX-based methods: Beyond usability and UI design | Human-centered AI experience design (see details below) |

**Table 1.** Comparison of cross-era characteristics of UX paradigms.

Figure 1 illustrates the evolution of UX paradigms over the past 40 years. In the first stage, UX methods focused on UI design and usability testing [3]. In the second stage, UX expanded to encompass broader concepts, such as end-to-end experience [4].

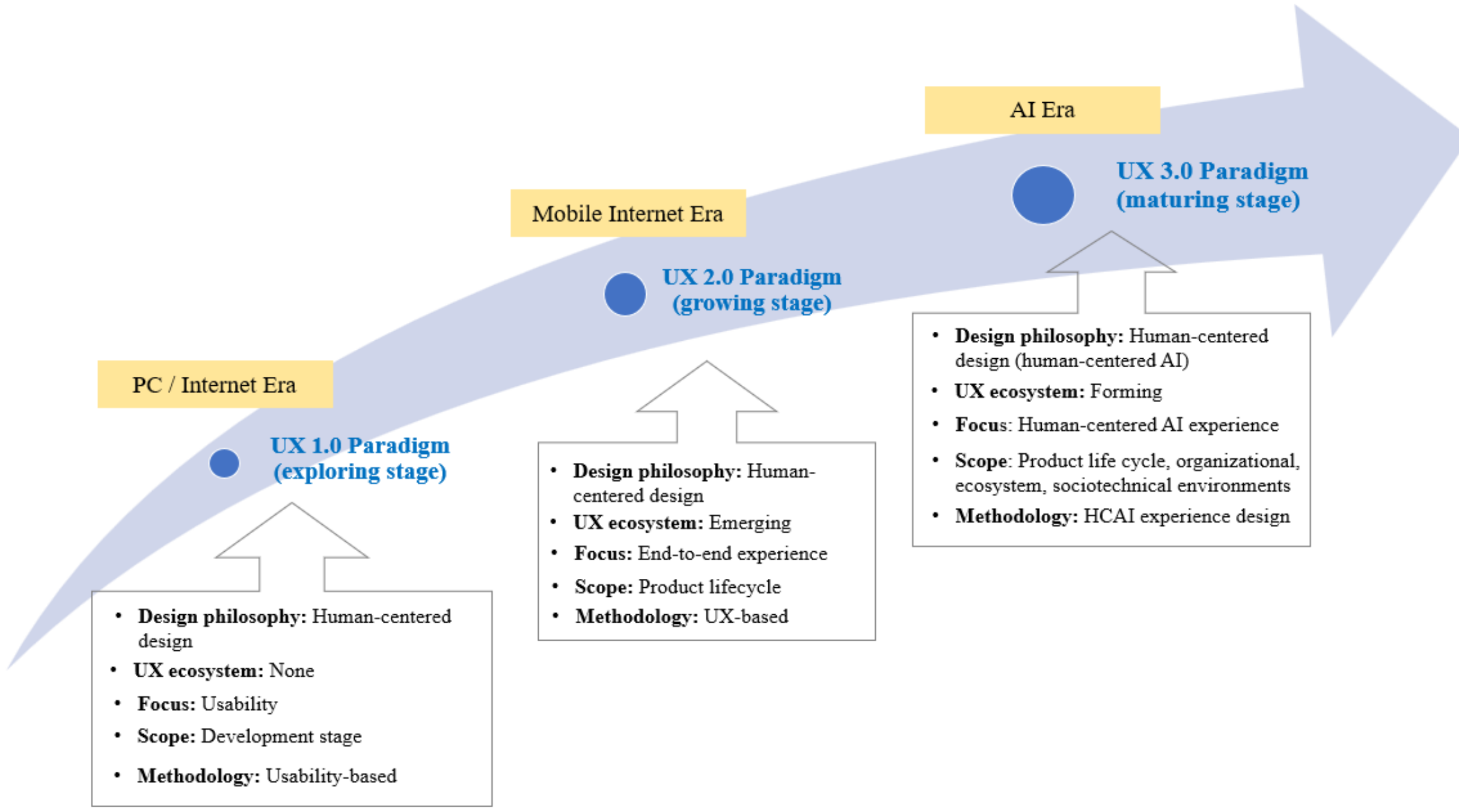

**Figure 1.** Evolution of UX paradigms across technological eras.

As we enter the third stage, the AI era, UX practice faces new challenges. While AI-based systems can benefit users, poorly designed or misused AI can lead to negative experiences and even harm. For example, thousands of AI-related accidents have been recorded, including self-driving-car crashes and algorithmic trading errors [5]. These failures often result from a technology-centered mindset when developing AI systems. In many cases, UX design activities are involved too late in the process, limiting their impact on and contribution to projects [6].

To address the unique challenges posed by AI technologies, an HCAI approach has been introduced [1,2], emphasizing the need to design AI systems that prioritize human values, needs, and experiences. Central to this vision is UX design, which plays a crucial role in making AI systems understandable, controllable, trustworthy, ethically aligned, and user-friendly. Traditional UX design has been practiced for non-AI systems and is insufficient for the complexity of AI-based systems. Therefore, a new UX paradigm must be developed to enable the realization of truly human-centered AI.

**Human-Centered AI Experiences and New Methodological Needs**

HCAI experiences refer to the positive experiences individuals gain when interacting with AI systems, extending beyond traditional experiences with non-AI technologies. Guided by the principles that we put together [7], HCAI experiences emphasize empowering users for greater productivity, ensuring human control over AI, aligning AI design with human values and needs, and fostering AI systems that are explainable, collaborative, safe, trustworthy, and easy to use. HCAI experiences can be classified as one of the following four key types, driving corresponding innovative UX design methodology.

***Ecosystem-based experience.*** At present, UX is no longer limited to its interactions with the UI of individual products at a time, as observed in UX 1.0 and UX 2.0; it now depends on its broader interactions beyond isolated instances. Accordingly, experience is increasingly shaped by engagement across diverse AI (e.g., multiple AI agents) and non-AI technologies as well as sociotechnical ecosystems.

*Across AI ecosystems*. UX is expanding beyond individual systems to encompass entire ecosystems, including mobile, wearable, cloud, and desktop platforms. Furthermore, AI-based ecosystems with multiple AI-agent systems, such as those in intelligent transportation and healthcare, are emerging. UX design must ensure that users have a seamless and consistent experience through integrated interactions across multiple AI systems within ecosystems.

*Across the entire AI life cycle.* UX spans the full AI life cycle, not just the UI design stage. It begins with predevelopment stages, such as branding and expectation management, continues through AI model training, testing, and deployment, and extends to postlaunch phases, including AI behavior monitoring, user feedback, model updates, user support, and end-of-life transitions. UX design must encompass managing user trust in AI behavior, addressing model drift, facilitating continuous learning and improvement, and incorporating ethical considerations throughout the AI life cycle.

*Across system architecture layers.* UX extends beyond the front-end UI of AI systems to encompass the entire system stack. This includes the middle layer, such as AI-driven business logic, decision models, AI algorithms, process automation, and the backend, which involves data pipelines, cloud infrastructure, and updates to AI models. A seamless and reliable experience is influenced by the integration across these layers, as well as other factors such as model transparency and data quality. UX design must optimize the integration across all these layers, through multidisciplinary collaboration with engineers and architects.

*Within a broad sociotechnical environment.* UX is affected by organizational, cultural, and social factors. UX design must consider how AI is integrated into the broader environment. This means designing AI systems that work well with both organizations and humans, such as ensuring trust, fairness, and cultural relevance, as well as optimizing work systems (e.g., roles, reskilling, workflows).

***Innovation-enabled experience.*** Innovation is about aligning user needs and real-life use with technology to make it useful and easy to use. Traditional innovation methods often focus heavily on technology, overlooking the role of UX, which contributes to high failure rates in product innovation. With products becoming increasingly similar, shorter development cycles, and rising user expectations, creating differentiated and meaningful experiences has become essential for successfully standing out in the AI era.

*Differentiated experience.* UX-driven innovation focuses on both solving current user pain points and uncovering unmet needs. By utilizing technologies such as AI and big data, UX design can identify common frustrations across similar products and create differentiated experiences. Also, AI predictive tools help discover potential user needs based on usage scenarios, enabling proactive experience design.

*AI-enabled experience.* AI technologies serve as enablers for innovative UX. Emerging interaction methods, such as voice input, can transform how users engage with AI systems, while existing technologies can be recombined to address challenges like addressing the motion problems of AR/VR through ~~using~~ multimodal interaction approaches. Real-time behavioral and contextual data, based on AI technology, also enable the delivery of personalized features and content.

*Enhanced experience by hybrid intelligence.* Effective human-AI intelligence requires dynamic functional allocation, where tasks are divided based on the strengths of humans and AI. By integrating their complementary capabilities, systems can support enhanced hybrid intelligence and an improved human controllability experience over AI. UX design needs to help build strong human-AI teams based on shared goals, mental models, and responsibilities, which enhances overall system performance and UX.

*End-to-end experience from a social perspective.* A sociotechnical approach enables holistic experience design that takes into account broader environmental factors through approaches such as innovative service design and process reengineering. This perspective ensures seamless end-to-end experiences by aligning technical capabilities with all touchpoints of users (e.g., Uber ride-share service) in social, cultural, and organizational contexts. UX design must address key interaction touchpoints between users and AI systems at various levels within such a broad environment.

***AI-enabled experience.*** AI-enabled experience redefines how users interact with AI systems. Users are no longer passive recipients, but active participants. AI transforms static user experiences into dynamic, context-aware, collaborative, and real-time enhanced experiences in the AI era.

*AI-enabled context-aware user experience.* AI enables users to have a more context-aware experience by allowing systems to understand user needs, intents, and behaviors through data-driven insights. By analyzing user behaviors, contextual activities, and multimodal inputs (such as gaze or facial expression), UX design can leverage AI to power the development of adaptive systems that provide more tailored and context-aware experiences.

*AI-enabled collaborative experience.* AI opens new frontiers in human-AI collaboration. With AI tools such as large language models and generative AI design platforms, users can engage in rapid, collaborative exploration of new ideas and interaction. UX design needs to facilitate dynamic collaboration by leveraging human and AI strengths, designing collaborative team-based UIs, and evaluating UX from the perspective of team performance and co-learning/coevolution.

*AI-enabled real-time enhanced experience.* Users benefit from AI systems that respond immediately and contextually. Whether through AI assistants, recommendation engines, or predictive personalization, AI systems deliver intuitive and tailored real-time experiences. UX design should support these capabilities to transform interaction into an ongoing, adaptive dialogue between users and AI systems.

***Human-AI interaction experience.*** Traditional HCI focused on human interaction with non-AI systems. Today, HCI is shifting toward interaction with AI systems, which offer richer experiences and create new demands for UX practice. AI possesses unique traits, such as autonomous and self-learning features, that make interactions more natural and powerful [1,2,7]. These features also create opportunities for UX design.

*User participatory experience.* As AI systems become increasingly autonomous and complex, their dynamic and sometimes unpredictable behavior directly influences UX, particularly when outputs appear biased, opaque, or misaligned with human needs. UX design should adopt a participatory approach that allows users to influence AI behavior through feedback, real-world testing, and iterative model refinement, enhancing users' sense of control and awareness of abnormal machine behaviors, ultimately leading to a more adaptive experience.

*Explainable AI for trust experience.* The “black box” nature of AI systems can reduce users' trust experience. UX design should focus on explainable AI that communicates how decisions are made. This includes using clear visualizations, transparent interface models, and explanation strategies grounded in psychological theories. These approaches enable users to understand, trust, and interact effectively with AI systems.

*Intelligent user experience.* AI-based intelligent UIs enable both humans and AI systems to initiate interactions based on user intent and context, transcending simple input methods. These interfaces enable more natural, social, and emotionally aware communication. UX design needs to incorporate techniques such as user intent and emotional recognition, LLM-based-prompt UI design, and socially responsive interaction models to create more intuitive experiences.

*Ethically aligned experience.* As AI becomes more integrated into everyday life, users demand more than just usability; they expect fairness, privacy, transparency, and control. UX design must address these ethical concerns by adopting sociotechnical and interdisciplinary design approaches. This includes ensuring that AI systems support user empowerment, human controllability, responsible decision making, and skill growth, all of which are aligned with human values and broader societal impacts.

**The UX 3.0 Paradigm Framework**

People have started to explore early ideas for UX paradigms suited to the AI era from a methodological perspective. Some are using AI and big data to analyze real-time user behavior and create personalized experiences. However, there is still no complete, systematic UX paradigm to support UX design in delivering HCAI experiences for AI systems.

As discussed earlier, UX 1.0 and 2.0 paradigms are typically applied to non-AI systems and often fall short when extended to AI systems. While AI introduces new challenges for UX design, it also offers an opportunity to evolve UX paradigms. To address these needs, we propose a conceptual framework for the UX 3.0 paradigm tailored to the AI era (Figure 2).

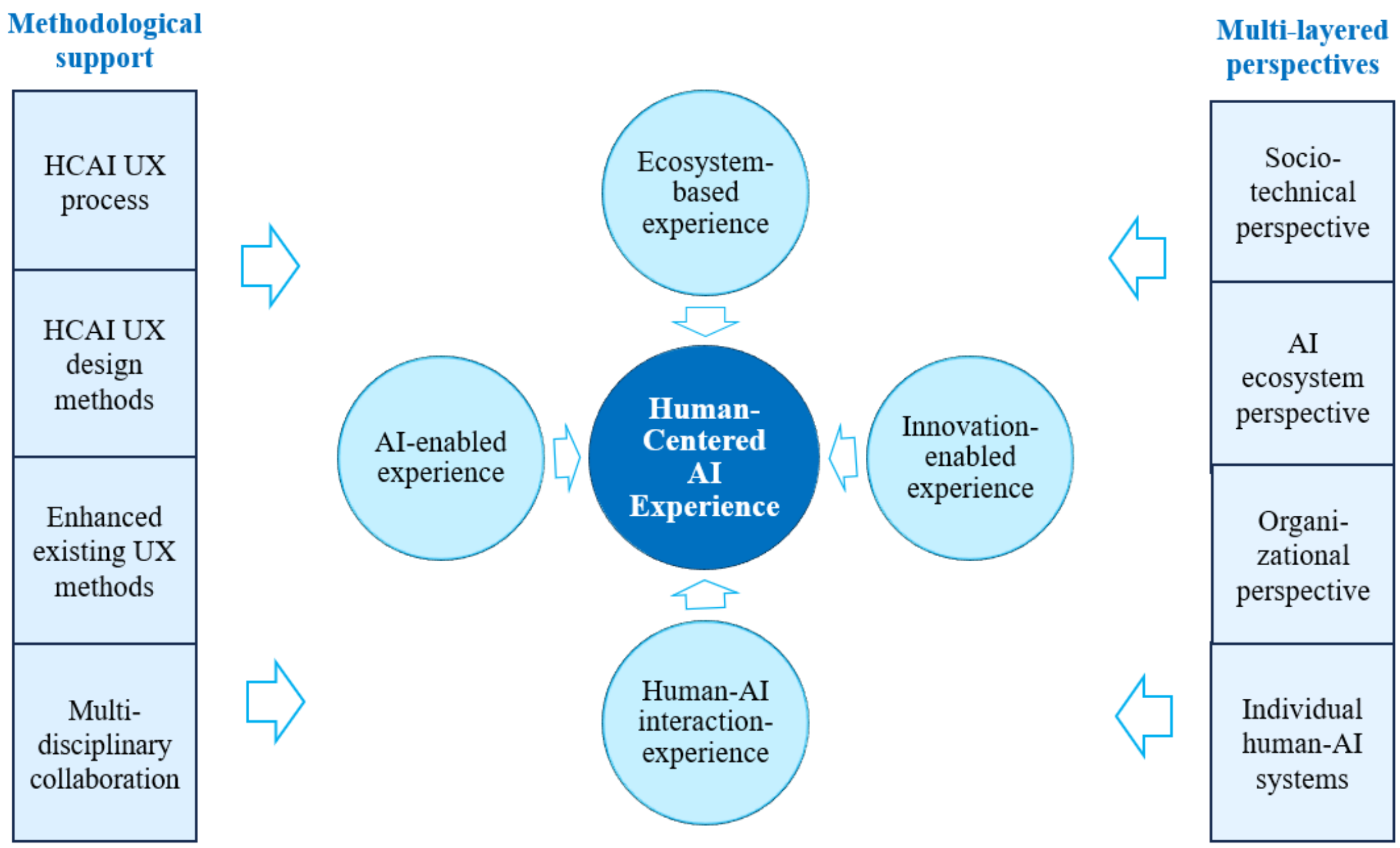

**Figure 2.** A conceptual framework of the UX 3.0 paradigm in the AI era.

The UX 3.0 paradigm framework aims to create human-centered AI experiences in the AI era. It responds to the new needs brought about by AI and aims to guide UX practice in the AI era.

*Emerging HCAI experiences.* Emerging HCAI experiences shape the UX 3.0 paradigm, calling for innovative and effective UX practice, including new design thinking, scope, methodology, and perspectives. The ecosystem-based experience expands UX design from individual products to connected AI systems from an ecosystem perspective. The innovation-enabled experience requires UX design to address product experiences through experience-driven innovation. The AI-enabled experience encourages UX design to integrate AI into user activities, providing a real-time experience for users. The human-AI interaction experience transforms UX design to focus on effective interactions between humans and AI.

*Methodological support.* The left column of the framework provides methodological support, outlining the key components needed to guide UX practice in the AI era. It includes an HCAI-based process to ensure that UX is integrated early and consistently into the entire AI life cycle. Innovative HCAI experience design methods and tools offer specific techniques, such as human-AI codesign, human-centered machine learning, and human-centered explainable AI, to address the complexity of HCAI experiences. Along with enhanced existing UX methods and multidisciplinary environments, these components form a comprehensive foundation for applying the UX 3.0 paradigm.

*Multilayered perspectives.* Different from UX 1.0 and 2.0, which focused on a siloed perspective, the right column of the framework presents multilayered perspectives that reflect the broader contexts in which UX design is practiced in the AI era. Beyond human-AI interaction with individual AI systems, the UX 3.0 paradigm extends the practice to organizational settings, where workers' experiences are influenced by work system redesign, including redefined roles, workflows, and skill sets. The perspective of AI ecosystems encourages UX design to address interactions across multiple AI systems. The sociotechnical perspective requires UX design to consider the joint optimization between AI technical and nontechnical subsystems, such as cultural, ethical, organizational, and social factors that shape UX outcomes in the real world.

**Implications and Future Work**

The UX 3.0 paradigm framework promotes a comprehensive approach to UX practice by emphasizing system-wide thinking that incorporates broad perspectives. It addresses emerging experiences in the AI era by focusing on the unique characteristics of AI technologies and the evolving expectations of users. UX 3.0 also provides actionable guidance for building human-centered AI experiences in AI systems. In addition, it calls for future work across design thinking, methodology, education, industry practice, and interdisciplinary collaboration.

*Evolving UX mindsets and practices.* UX design must shift from traditional usability-centered approaches to a more holistic, strategic, and ecosystem- and socio-aware mindset. Designing for human-centered AI experiences requires not only user-friendly interfaces but also alignment with AI behavior, ethics, explainability, controllability, and system-wide integration. This transformation requires new design thinking and practices.

*Advancing UX methodology.* The UX 3.0 paradigm framework encourages the development of specific tools, methods, and evaluation criteria for HCAI experiences. Future work should focus on advancing and validating these UX methods [7]. These efforts will make UX 3.0 more actionable across the full AI product life cycle, from data collection and model training to deployment, adaptation, AI behavior monitoring, and end-of-life management.

*Strengthening multidisciplinary collaboration.* To effectively implement UX 3.0, collaboration with professionals such as AI engineers, human factors scientists, data scientists, ethicists, and domain experts is critical. Future work should also explore a human-centered joint optimization of AI technical systems with sociotechnical factors, including culture, organization, and ethics. This calls for integrated methodologies through multidisciplinary collaboration.

*Cultivating UX talent.* The AI era demands a new generation of UX professionals who are fluent in both design and AI. College UX curricula should include courses on human-AI interaction, machine behavior, explainable AI, and ethical design. Graduate programs and research initiatives must also cultivate advanced skills in AI-driven UX methods and interdisciplinary research.

*Fostering organizational adoption.* Organizations should embed UX 3.0 principles into their product strategies, development processes, and AI governance models. This includes creating cross-functional UX/AI teams, integrating UX into early AI design stages, and fostering a culture that values human-centered innovation. Future work should explore frameworks for assessing UX maturity in AI ecosystems and measuring the effectiveness of human-centered AI experiences.

## Conclusion

Over the past 40 years, UX has developed in stages, driven by emerging technologies, changing user needs, and evolving practices. As UX enters the AI era, it faces new challenges and demands. I introduce the UX 3.0 paradigm, which aims to address the new challenges and offer solutions, urging professionals to recognize that current UX practices are no longer enough and must be improved. Advancing UX practice in the AI era will ultimately lead to human-centered AI systems that benefit both humans and society.

**Wei Xu** is a Professor of human factors/HCI at Zhejiang University in China. He is an elected fellow of the International Ergonomics Association, the Human Factors and Ergonomics Society, and the Association for Psychological Science. His research interests include human-centered AI, human-AI interaction, and aviation human factors. weixu6@yahoo.com